\documentclass[aps,prl,twocolumn,showpacs,groupedaddress]{revtex4}% review and submission
\usepackage{graphicx}  % needed for figures
\usepackage{dcolumn}   % needed for some tables
\usepackage{bm}        % for math
\usepackage{amssymb}   % for math
\usepackage{verbatim}

\usepackage{amsmath}
\usepackage{amsfonts}

\newcommand{\ket}[1]{|#1\rangle}

\begin{document}
\title{1/f Flux Noise in Josephson Phase Qubits}
\author{Radoslaw C.~Bialczak$^1$}
\author{R.~McDermott$^{2}$}
\author{M. Ansmann$^1$}
\author{M. Hofheinz$^1$}
\author{N. Katz$^1$}
\author{Erik~Lucero$^1$}
\author{Matthew Neeley$^1$}
\author{A. D. O'Connell$^1$}
\author{H.~Wang$^1$}
\author{A. N. Cleland$^1$}
\author{John M. Martinis$^{1*}$}

\affiliation{$^{1}$Department of Physics, University of California,
Santa Barbara, California 93106, USA}

\affiliation{$^{2}$Department of Physics, University of
Wisconsin-Madison, Madison, Wisconsin 53706}
\email{martinis@physics.ucsb.edu}

\date{\today}

\begin{abstract}
We present a new method to measure $1/f$ noise in Josephson quantum
bits (qubits) that yields low-frequency spectra below 1 \textrm{Hz}.  
Comparison of noise taken at positive and negative bias of a phase
qubit shows the dominant noise source to be flux noise and not
junction critical-current noise, with a magnitude similar to that
measured previously in other systems.  Theoretical calculations show that
the level of flux noise is not compatible with the standard model of
noise from two-level state defects in the surface oxides of the films.  
\end{abstract}

\maketitle

Superconducting integrated circuits are a leading candidate for
scalable quantum information processing (QIP)$\!$~\cite{general}.$\,\,\,$Quantum bits (qubits)
based on Josephson junctions have already achieved several key
milestones, including single and coupled qubit state tomography~\cite{matthiasPRL,matthias}.  
Moreover, the dominant mechanism for energy relaxation is becoming
understood~\cite{johnPRL}, and steady improvements can be expected in the coming
years.  However, to realize the full potential of Josephson junctions
for QIP, it will be necessary to extend qubit dephasing times.  
Present dephasing times are in the 100's of ns range; the short
coherence places a strict limit on the number of gate operations
which can be implemented, and represents a significant obstacle to
scaling up.  Dephasing is produced by low-frequency fluctuations in
the qubit energy.  In the case of the Josepshon flux qubit and the
flux-biased Josephson phase qubit, these fluctuations are believed
to arise from a magnetic flux noise applied to the qubit loop, with
a spectral density that scales inversely with frequency $(1/f)$.
Moreover, the magnitude of the flux noise inferred from qubit Ramsey
fringe experiments is of the order of several $\mu\Phi_0/\sqrt{\textrm{Hz}}$
for both three-junction flux qubits and phase qubits, despite a difference in loop inductance of almost two
orders of magnitude~\cite{nakamura}.

Low-frequency noise in superconducting circuits has been studied for
decades in the context of amplifiers based on the Superconducting
QUantum Interference Devices (SQUIDs)~\cite{kochOld,wellstood}.  More that 20 years ago in a series of
experiments on SQUIDs cooled to millikelvin temperatures,
researchers found that the devices displayed a flux noise with a
power spectrum which scaled like $1/f^\alpha$ at low frequencies,
where $\alpha$ lies in the range from 0.6 to 1.  The magnitude of the
noise was seen to be only weakly dependent on a wide range of device
parameters such as SQUID loop inductance, geometry, material, etc.,
with a canonical value at $1\ \textrm{Hz}$ of about $2\
\mu\Phi_0/\sqrt{\textrm{Hz}}$.  The origin of the excess low temperature flux
noise in these experiments was never understood, and the issue has
lain dormant for almost two decades.  Now it seems clear that the
excess low-temperature noise of these SQUIDs is intimately connected
to the measured dephasing times of superconducting qubits~\citep{nakamura,johnPRB,kochNew}.  

In this Letter we present the results of a novel measurement in a
Josephson phase qubit that uses the resonant response of the qubit to directly
measure the spectrum of low-frequency noise.  This general method can be used for any qubit system.  By alternating the
sense of the qubit bias, we show that the noise is predominantly
flux-like, as opposed to a critical-current noise.  This experiment
is the first to directly connect flux noise in superconducting
qubits to previous measurements in SQUID devices.  Additionally, we
present the results of calculations of flux noise from paramagnetic defects
in the native oxides of the superconductors, and show that the
measured flux noise is not compatible with the standard model of
two-level state (TLS) defects.  

\begin{figure}[b]
\includegraphics[width=.47\textwidth]{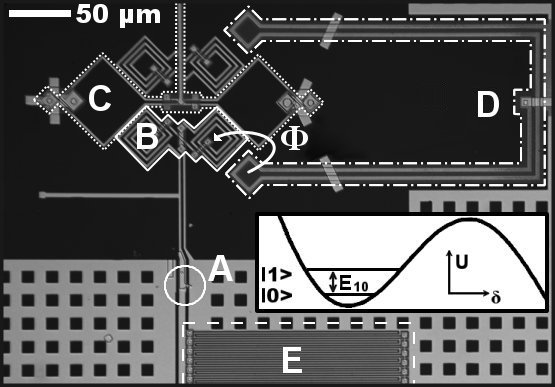}
\vspace*{-0.15in} \caption{Photomicrograph of Al-based qubit fabricated on a sapphire substrate using a $\textrm{SiN}_x$ dielectric for crossover wiring.  (A) Josephson junction with area $A_J \sim 2\
\mu\textrm{m}^2$ and critical current $I_0=1.9\: \mu\textrm{A}$.  (B) Qubit inductor with inductance $L=800\ \textrm{pH}$.  (C) Readout SQUID.  (D) Qubit flux bias.  (E) Qubit shunt capacitor with $C=1\ \textrm{pF}$.  (inset) Qubit potential energy $U$ as a function of the superconducting phase $\delta$ across the qubit Josephson junction.  }
\label{fig:figure1}\end{figure}

\begin{figure*}
\includegraphics[width=1\textwidth ]{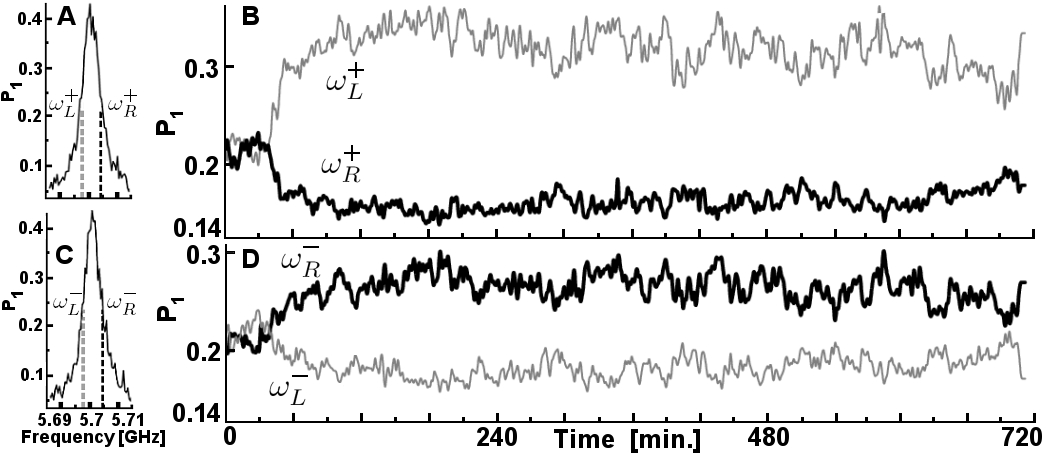}
\vspace*{-0.27in} \caption{(A) Qubit response curve for the probability $P_1$ of the qubit in the $\ket{1}$ state versus microwave excitation frequency at the positive flux bias $\Phi_+$.  (B) The time evolution of $P_1$ is measured for $\Phi_+$ at the two frequencies $\omega_L^+$ and $\omega_R^+$, which are shown in (A).  (C) and (D) are the same as (A) and (B), respectively, but are for $\Phi_-$.  In (B) and (D), data was taken at approximately 800 samples per second.  The correlation in $\omega_L^+$ and $\omega_R^-$ indicates flux noise.}\vspace*{-0.15in} \label{fig:figure2}
\end{figure*}

A photomicrograph of our device is
shown in Fig.~\ref{fig:figure1}; a more detailed discussion of its
operation is given elsewhere~\citep{nadav}.  A current bias $I=\Phi/L$ is applied to
the Josephson junction via a flux $\Phi$ threading an
inductor $L$ placed across the junction.  The bias current is set
slightly below the critical current of the junction $I_0$ so that
the system can be well modeled by a cubic potential (see
inset).  The two lowest quantum states in
this potential well are labeled as qubit states $\ket{0}$ and
$\ket{1}$, and have an energy difference $E_{10}$ that can be
tuned with bias.  Transitions between $\ket{0}$ and $\ket{1}$ are driven
by applying microwaves at a frequency $\omega_{10} / 2\pi =
E_{10}/h \sim 6\ \textrm{GHz}$.  The qubit state is measured by
applying a fast bias pulse to lower the potential barrier, forcing
only the $\ket{1}$ state to tunnel out of the well~\citep{ken}.  

The transition frequency $\omega_{10}$ is given by
\begin{equation}
\omega_{10} \simeq \omega_{p0}  \left( 1 -|I|/I_0\right)^{1/4} \
,\label{eqnn1}
\end{equation}
where $\omega_{p0} = 2^{1/4}(2\pi I_0/C\Phi_0)^{1/2}$.  Low
frequency fluctuations in the current bias $I$ and critical
current $I_0$ produce fluctuations in the transition frequency
primarily from the second term.  The qubit can be operated at both
positive and negative current bias.  A positive fluctuation in $I_0$
gives an increase in $\omega_{10}$ at both positive and negative
current bias, a symmetric change.  A fluctuation in $I$, however,
gives an asymmetric change.  Therefore, a spectroscopic measurement of the
transition frequency at positive and negative bias currents provides
a clear differentiation between these two different noise sources.  
Moreover, the steep response of the resonance allows for a reasonably
sensitive measurement of the fluctuation magnitude.  

The experiment is performed by choosing positive and negative
current biases close to the critical current, corresponding to flux
biases $\Phi_{+}$ and $\Phi_{-}$, that have approximately equal
transition frequencies.  A spectroscopic measurement is then
performed by applying a long $2\ \mu\textrm{s}$ microwave pulse and
measuring the probability $P_1$ of the occupation of the $\ket{1}$
state.  The amplitude of the microwave excitation is chosen so that
$P_1 \lesssim 0.4$ at peak response to prevent significant power broadening of the qubit response.  
Qubit response curves for positive and negative biases are
shown in Figs.~\ref{fig:figure2}A and~\ref{fig:figure2}C.  

As shown by Eq.\,(\!\!~\ref{eqnn1}), fluctuations in bias and critical current will cause these resonance
curves to shift.  The probability $P_1$ is most sensitive to qubit bias at the
half maximum points of the resonance curves, labeled as
frequencies $\omega_L^-$, $\omega_R^-$ and $\omega_L^+$,
$\omega_R^+$ in Figs.~\ref{fig:figure2}A and~\ref{fig:figure2}C
respectively.  A plot of $P_1$ versus time is shown for these four
frequencies in Figs.~\ref{fig:figure2}B and~\ref{fig:figure2}D.  An
anti-correlated change in $P_1$ within the data pairs ($\omega_L^-$,
$\omega_R^-$) and ($\omega_L^+$, $\omega_R^+$) is expected, and
represents a systematic check of the measurement method.  The small
deviations from anti-correlation are due to other influences, such as
fluctuations of resonant TLS defects~\citep{ken} which affect the
measurement probability pairs in a correlated manner.  

\begin{figure}[b]
\includegraphics[width=.49\textwidth]{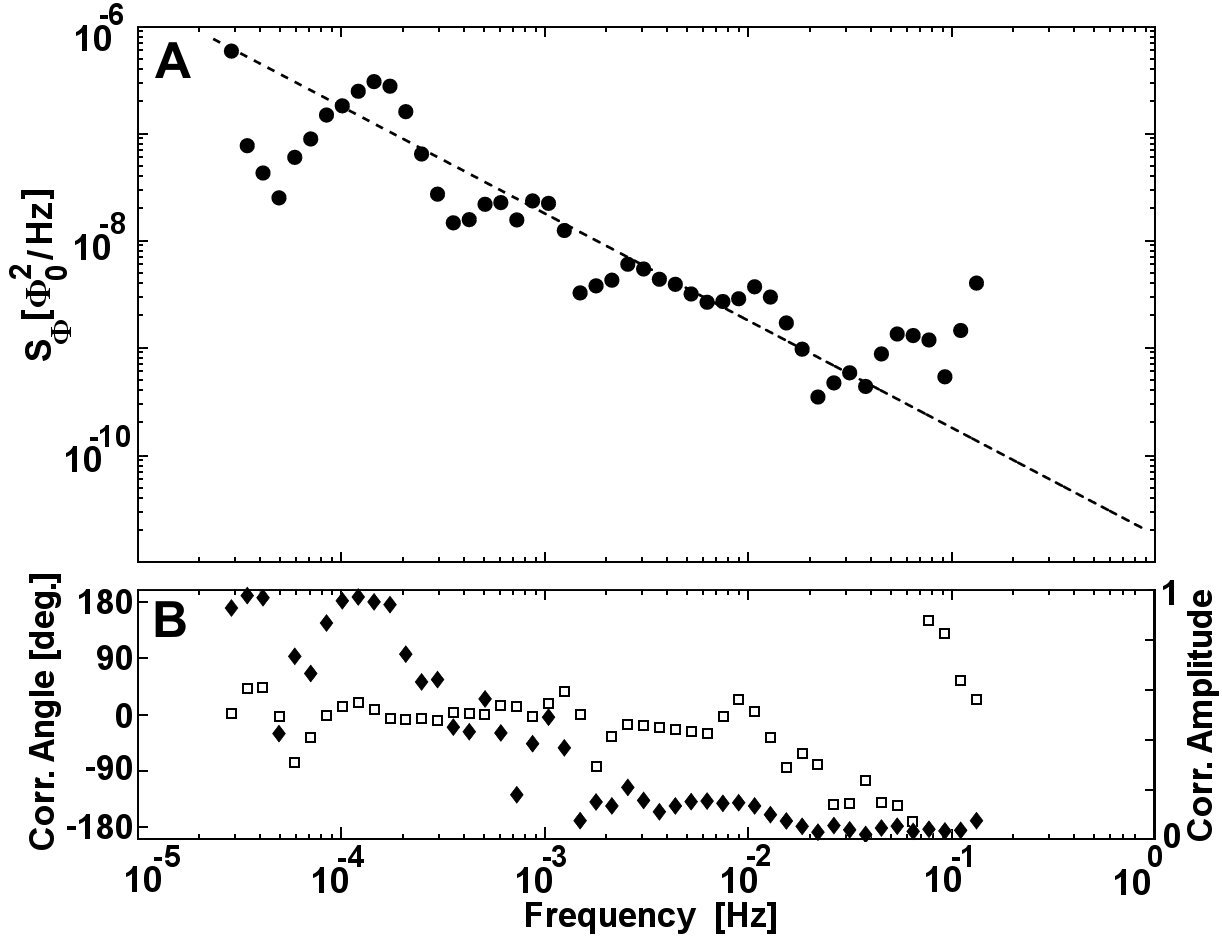}
\vspace*{-0.17in}\caption{(A) Cross-correlated noise power spectrum of data taken from Figs.~\ref{fig:figure2}(B) and (D).  The line for an ideal $1/f$ spectrum is shown for reference.  (B) Correlation amplitude (diamonds) and phase angle (squares) of cross-correlated data in (A).}
\label{fig:figure3}
\end{figure}

The data at $\omega_L^+$ and $\omega_R^-$ give a symmetric
correlation of $P_1$ with time, as can be seen from the traces
in Figs.~\ref{fig:figure2}B and~\ref{fig:figure2}D.  This shows that the dominant low-frequency
noise for the qubit is a flux noise.  The relation between $P_1$
and flux is calibrated by measuring $P_1$ while sweeping the qubit flux bias
for each of the four frequencies.  The
flux noise data measured at $\omega_L$ and $\omega_R$ are averaged for
positive and negative bias, then Fourier transformed,
cross-correlated, and averaged over frequency to obtain the cross-correlated flux
noise spectrum plotted in Fig.~\ref{fig:figure3}A.  We note that white
noise from the measurement process is automatically subtracted in
this cross-correlation analysis.  The noise has a $1/f^{\alpha}$
spectrum with $\alpha = 0.95$ and extrapolates to a flux noise at
$1\ \textrm{Hz}$ of $4\ \mu \Phi_0 / \sqrt\textrm{Hz}$.  This
magnitude is comparable with previous measurements of $1/f$
flux noise in superconducting devices~\citep{kochOld,wellstood}.    

In Fig.~\ref{fig:figure3}B we plot the correlation amplitude and phase angle of the cross-correlated flux noise
spectrum.  The correlation angle of zero indicates asymmetric (flux-like)
noise.  At the lowest frequencies, the contribution from measurement noise is small.  The saturation of the
correlation amplitude at a value slightly less than unity may indicate a small contribution
from junction critical-current noise.  Taking the contribution from critical-current noise to be $\sim 5 \%$,
we find $S_{I_0}(1\ \textrm{Hz}) = 0.05\ S_\Phi(1\ \textrm{Hz})/L^2 = 1.4\times 10^{-12} I_0^2/\textrm{Hz}$ for our $2\ \mu\textrm{m}^2$ area junction.  
This value is compatible with previous experiments~\cite{wellstood}, giving a critical-current noise at $20\ \textrm{mK}$ that is a factor of 36 lower than predicted for $4.2\ \textrm{K}$~\cite{savo}.  

We note that flux noise produces dephasing of the qubit state, as can be measured directly in a Ramsey fringe experiment.  
The magnitude of our noise is within a factor of two of that required to explain our qubit dephasing times of around 200 ns~\citep{johnPRB}.  

In what follows, we examine the possibility that the flux noise is due
to magnetic TLS defects in the native oxides of the superconducting films, as was recently proposed in Ref.~\citep{kochNew}.  
The standard TLS model~\citep{standardTLS} describes an ensemble of defects, each with two
microscopic configurational states $\ket{L}$ and $\ket{R}$ that have a
two-state Hamiltonian with diagonal matrix elements $\pm\Delta/2$
and off-diagonal elements $\Delta_0/2$ due to tunneling.  The
eigenstates are given by
$\ket{g}=\sin(\theta/2)\ket{L}+\cos(\theta/2)\ket{R}$ and
$\ket{e}=\cos(\theta/2)\ket{L}-\sin(\theta/2)\ket{R}$, where
$\theta=\arctan(\Delta_0/\Delta)$.  The difference in energy of the
two states is $E=\sqrt{\Delta^2+\Delta_0^2}$.  The defects are
assumed to have a constant distribution of energies $\Delta$, but
a log-uniform distribution in $\Delta_0$ because tunneling is exponentially
dependent on parameters.  Upon changing variables to
$(E,\sin\theta)$, the joint distribution is given by $d^2N=\rho\ dE\
d\sin\theta/\sin\theta \cos\theta$, where $\rho$ is a materials
constant describing the defect density of states.  Dipole radiation of the TLS via phonons gives a relaxation rate determined by the matrix element $\sin\theta$, yielding
$\Gamma_1 = \Gamma_1^\textrm{max}\sin^2\theta$.  The resulting log-uniform distribution of
$\Gamma_1$ produces a $1/f$ noise spectrum.  

To estimate the magnitude of the flux noise from magnetic TLS defects, we consider
a TLS magnetic moment equal to the Bohr magneton $\mu_B$, and further assume that fluctuation of the TLS will
completely randomize this magnetic moment (we discuss the validity of this assumption below).  
Following the analysis of Ref.~\citep{shnirman}, one can show that the low-frequency spectral density of the TLS
magnetic moment per unit volume is given by
\begin{align}
S_m(\omega/2\pi) &\simeq 4kT \mu_B^2 \rho \int_0^{\Gamma_1^\textrm{max}}
\frac{d\Gamma_1}{2\Gamma_1}
 \frac{2\Gamma_1}{\Gamma_1^2+\omega^2} \\
& \simeq  \frac{kT \mu_B^2 \rho}{\omega/2\pi} \ .
\label{TLSspectral}
\end{align}

In order to connect the above expression to the measured flux noise, we need to know how each TLS couples magnetically to the SQUID.  Analytical expressions for flux noise may be calculated using reciprocity: the magnetic flux from a spin of moment $m$ is given by $(B\cdot m)/I$, where $B$ is
the magnetic field at the spin produced by a test current $I$ in the SQUID loop.  
We consider two idealized SQUID geometries that are amenable to
analytical treatment.  First, we consider a thin wire of
diameter $D$ in a circular loop of radius $R$ with $R \gg D$.   
We find that the mean-square flux induced in the SQUID by the TLS defects is given by
\begin{equation}
\langle \Phi^2 \rangle = \frac{2 \mu_0^2}{3} \mu_B^2
\sigma \frac{R}{D} \ , \label{fluxsimple}
\end{equation}
where $\sigma$ is the density of TLS surface defects on
the superconducting wire.  A factor $1/3$ arises from a random
angular distribution of the TLS magnetic moments.  For the more realistic geometry
of a thin-film superconductor of width $W$ and thickness $b$ in a
circular loop of radius $R$ with $R \gg W \gg b$, the surface
currents $J(x)$ at position $-W/2+\lambda < x < W/2-\lambda$ are
proportional to $[1-(2x/W)^2]^{-1/2}$, where $\lambda$ is the
penetration depth~\citep{vanduzer}.  The currents fall away exponentially to zero at
the edges $\pm W/2$.  With the surface magnetic field being proportional
to the surface current density, the mean-square flux coupled to the SQUID is
calculated as follows:
\begin{align}
\langle \Phi^2 \rangle &= (\pi/6) \mu_0^2\mu_B^2\sigma R
\frac{\int dx\ J^2(x)}{[\int dx\ J(x)]^2} \\ &\simeq \frac{2
\mu_0^2}{3}\mu_B^2\sigma\frac{R}{W} \
\big[\frac{\ln(2bW/\lambda^2)}{2\pi} + 0.27\big] \ .
\label{eqthinfilm}
\end{align}

\noindent The logarithmic term changes the prediction of Eq.\,(\ref{fluxsimple}) by a factor $\sim1.8$, with a reasonable fraction of the noise arising from fluctuators within a few penetration depths near the edges of the film.  

The major geometric dependence of the noise comes from the ratio $R/W$, the loop radius to width, with only a logarithmic dependence on the overall scale~\citep{footnote1}.  This feature of the model is compatible with the observation that the flux noise of $\mu\textrm{m}$-sized flux qubits is similar to that found for our $200\ \mu\textrm{m}$ scale qubit, as the geometric ratio $R/W$ is similar for these devices.  

The critical parameter determining the magnitude of the noise is the surface
density of defect states $\sigma=\rho t$, where $t$ is the thickness of the
surface oxide on the superconducting film.  The TLS defect density in amorphous oxide
films can be extracted from measurements of the loss tangent of large-area
tunnel junctions~\citep{johnPRL}.  It is found that this defect density is compatible with bulk values obtained for a wide variety of amorphous oxides.  We therefore take as an estimate of the TLS surface density~\citep{johnPRL} $\rho t=1.0/\mu \textrm{m}^2(h \textrm{GHz})$, twice that measured
in tunnel junctions, to account for the thicker surface oxide $t\sim$ 2 nm.  
To calculate the spectral density of the flux noise, we substitute
$S_mt$ for $\mu_B^2 \sigma$ in Eq.\,(\ref{eqthinfilm}).  
Using the parameters $R/W=10$ and $T$ = 100 mK, we compute a flux noise spectral density
$S_\Phi(1\ \textrm{Hz}) = 1.1\times 10^{-3} (\mu\Phi_0)^2/\textrm{Hz}$, about $4$ orders of magnitude
smaller than the measured flux noise.  

Although we have not explicitly considered the noise contribution
from other surfaces away from the superconductor or dielectrics in
crossover wiring, these small volumes cannot  compensate
for the large discrepancy between the measured and calculated noise.  The substrates are not likely
candidates since they are typically crystalline,
and therefore have very low defect densities.  In addition, defects at a $\textrm{Si/SiO}_\textrm{x}$ interface cannot account for the measured noise since our devices were made on sapphire substrates.  

Moreover, we note that the assumption that TLS fluctuation randomizes the defect magnetic moment is
highly questionable because TLS defects in typical oxides are not considered to be magnetic.  The above density of magnetic defect states is probably a gross overestimate, further exacerbating the discrepancy between the measured flux noise and the noise calculated from TLS defect states.  

If spin noise is responsible for flux noise, we conclude that it
must arise from a surface defect mechanism that is very different
than that described by the standard TLS model, as it must have a defect/atomic-bond
ratio that is about $10^4$ times larger than for bulk TLS defects.  
Such a model would predict that specific heat measurements for
amorphous materials would be dominated by surface states once the
thickness of structures is less than about $1 - 10 \ \mu\textrm{m}$.  
Koch \textit{et al.} have suggested~\citep{kochNew} surface electronic states as a
possible candidate; unfortunately, the density of these defects has been estimated
only at room temperature.  A possible new mechanism has been
proposed based on tunneling of conduction electrons into surface
states~\citep{faoro}.  

In conclusion, we have demonstrated a new measurement of $1/f$ flux
noise in superconducting qubits, which allows us to distinguish
between flux and critical-current fluctuations.  The magnitude of
the measured noise is in good agreement with previous experiments,
even though device parameters greatly differ.  We have also
theoretically considered a spin-noise mechanism arising from
fluctuating TLS.  With the predicted magnitude in disagreement by over 4
orders of magnitude, we conclude that any model for spin noise must
arise from a new mechanism based on a high density of defects.  

We would like to thank C. Yu, L. Faoro, and L. Ioffe for discussions on magnetic TLS defects.  
Devices were made at the UCSB and Cornell NanoScale Facilities, a part of the NSF-funded
National Nanotechnology Infrastructure Network.  This work
was supported by Disruptive Technology Office under grant
W911NF-04-1-0204 and by NSF under grant CCF-0507227.

\end{document}